\documentclass[aps,prl,twocolumn,superscriptaddress,preprintnumbers,showpacs,showkeywords,floatfix]{revtex4-1}

\usepackage{graphicx}
\usepackage{dcolumn}
\usepackage{bm}
\usepackage{amsmath}
\usepackage{hyperref}
\usepackage{xcolor}

\begin{document}

\preprint{ADP-12-34/T801}

\title{Low-lying Odd-parity States of the Nucleon in Lattice QCD}

\author{M. Selim Mahbub} 
\author{Waseem Kamleh}
\author{Derek B. Leinweber}
\author{Peter J. Moran}
\author{Anthony G. Williams}
\affiliation{Special Research Centre for the Subatomic Structure of
  Matter, School of Chemistry \& Physics, University of Adelaide, SA
  5005, Australia}

\collaboration{CSSM Lattice Collaboration}

\begin{abstract}
The world's first examination of the odd-parity nucleon spectrum at
light quark masses in $2+1$ flavor lattice QCD is
presented. Configurations generated by the PACS-CS collaboration and
made available through the ILDG are used, with the lightest pion mass
at $156$ MeV. A novel method for tracking the individual energy
eigenstates as the quark mass changes is introduced.  The success of
this approach reveals the flow of the states towards the physical
masses.  Using the correlation matrix method, the two lowest-energy
states revealed are found to be in accord with the physical spectrum
of Nature.
\end{abstract}



\maketitle

Lattice QCD is the only currently known {\it ab-initio} or
first-principles approach to the fundamental quantum field theory
governing the properties of hadrons, Quantum Chromodynamics
(QCD). While the ground-state hadron spectrum of QCD is well
understood, a determination of the excited state
energy spectrum is in the process of being revealed.  Ultimately, the
results can be compared with the existing experimental data and
provide predictions and motivations for future experiments.

Hadron spectroscopy is dependent on the rich dynamics of the strong
interaction.  For example, the experimentally observed mass of the
first positive-parity excitation of the nucleon, known as the Roper
resonance, $N{\frac{1}{2}}^{+}$ $(1440)$ ${\rm{P}}_{11}$, is
surprising low compared to the lowest-lying negative-parity partner,
$N{\frac{1}{2}}^{-}$ $(1535)$ ${\rm{S}}_{11}$.  This phenomenon is not
observed in constituent or valence quark models where the lowest-lying
odd-parity state occurs naturally {\it below} the first
$J^P={\frac{1}{2}}^{+}$ excitation.

Drawing on experimental results, we note that the Breit-Wigner width
of the $N^{-}(1535)$ state is $\approx 150$ MeV, approximately half
the width of the Roper $N^{+}(1440)$~\cite{Nakamura:2010zzi}.
Furthermore, the branching fraction $\Gamma(\pi N)/\Gamma$ for
$N^{-}(1535)$ is $2/3$ of the Roper.  Together, these factors indicate
a suppression of $1/3$ in the coupling of $\pi N$ to the $N^{-}(1535)$
state relative to the Roper.  Noting that the light $\pi N$ dressing
makes the most important self-energy contribution, it is anticipated
that the self-energy dressings of $\pi N$ for the $N^{-}(1535)$ will
be reduced to approximately $10\%$ of that for the Roper.  A
consequence of this is to suppress the finite-volume effects of the
lattice QCD simulation which can otherwise lead to large energy shifts
associated with the avoidance of energy-level crossings of the single
and multi-particle scattering states.  Similar arguments for the
$N^{-}(1650)$ suggest $\pi N$ self-energy contributions are suppressed
to the 25\% level.  Thus, it is interesting to directly compare the
results of our lattice QCD simulations with experiment and gain
insight on the quark mass dependence of these states.
While finite-volume effects are of residual interest in this
investigation, understanding the finite volume effects on these states
and linking them to the resonances of Nature is a long term program of
the lattice QCD community.

The experimentally observed nearly-degenerate $S_{11}$ $(1535)$ and
$(1650)$ states are in agreement with the simple quark-model
predictions based on $SU(6)$ symmetry. Therefore, looking at the
low-lying $N{\frac{1}{2}}^{-}$ energy states and their structure from
the first principles approach is potentially very revealing. Some
recent full QCD results can be seen in
Refs.~\cite{Bulava:2009jb,Bulava:2010yg,Engel:2010my,Edwards:2011jj,Mahbub:2010rm,Menadue:2011pd}.
Herein, it will be interesting to explore the physics associated with
the dynamical fermion loops of full QCD, this time at very light quark
masses.

The correlation functions for the $N{\frac{1}{2}}^{-}$ states are
short-lived compared to the lighter $N{\frac{1}{2}}^{+}$ ground state.  Thus
it is important to adopt a method that can isolate the effective-mass
plateaus at early Euclidean times.  The variational
method~\cite{Michael:1985ne,Luscher:1990ck} is the state-of-the-art
approach for achieving this in lattice hadron-spectroscopy
calculations and is adopted here.  Through a generalized eigenvalue
analysis of a matrix of correlation functions, the process enables one
to create highly optimized interpolating fields designed to excite a
single energy eigenstate of the QCD Hamiltonian.  The masses of the
energy states are then obtained through a standard effective-mass
analysis~\cite{Mahbub:2009aa} providing a robust approach for
extracting the energy states at early Euclidean times.

In this paper, we utilize the established approach of
Refs.~\cite{Mahbub:2010rm,Menadue:2011pd} to explore the low-lying
$N{\frac{1}{2}}^{-}$ energy states in full QCD.  In doing so, a novel
method has been developed to track the energy eigenstates from heavy
to light quark masses.  The method is particularly useful when the
energy-states are nearly degenerate.

The two-point correlation-function matrix for $\vec{p} =0 $ can be
written as
\begin{align}
G_{ij}^{\pm}(t) &= \sum_{\vec x}\, {\rm Tr}_{\rm sp}\, \{
\Gamma_{\pm}\, \langle\, \Omega\, \vert\, \chi_{i}(x)\,
\bar\chi_{j}(0)\, \vert\, \Omega\, \rangle\}, \\
          &=\sum_{\alpha}\, \lambda_{i}^{\alpha}\,
\bar\lambda_{j}^{\alpha}\, e^{-m_{\alpha}t},
\end{align}
where Dirac indices are implicit, $\lambda_{i}^{\alpha}$ and
$\bar\lambda_{j}^{\alpha}$ are the couplings of interpolators
$\chi_{i}$ and $\bar\chi_{j}$ at the sink and source respectively,
$\alpha$ enumerates the energy eigenstates with mass $m_{\alpha},$ and
$\Gamma_{\pm}=(\gamma_{0}\pm 1)/2$ projects the parity of the
eigenstates.  A linear superposition of interpolators
$\bar{\phi}^{\alpha}=\sum_{j}{\bar\chi}_{j}u_{j}^{\alpha}$ creating
state $\alpha$ provides the relationship
\begin{align}
G_{ij}(t_{0}+\triangle t)\, u_{j}^{\alpha} & = e^{-m_{\alpha}\triangle
  t}\, G_{ij}(t_{0})\, u_{j}^{\alpha}  \, ,
 \label{eq:recurrence_relation}
\end{align}  
from which right and left eigenvalue equations are obtained
\begin{align}
[(G(t_{0}))^{-1}\, G(t_{0}+\triangle t)]_{ij}\, u^{\alpha}_{j} & = c^{\alpha}\, u^{\alpha}_{i}, \label{eq:right_evalue_eq}\\
v^{\alpha}_{i}\, [G(t_{0}+\triangle t)\, (G(t_{0}))^{-1}]_{ij} & = c^{\alpha}v^{\alpha}_{j},\label{eq:left_evalue_eq}
\end{align} 
with $c^{\alpha}=e^{-m_{\alpha}\triangle t}$.
The vectors $u_{j}^{\alpha}$ and $v_{i}^{\alpha}$ diagonalize the
correlation matrix at time $t_{0}$ and $t_{0}+\triangle t$ making the
projected correlation matrix,
$v_{i}^{\alpha}G_{ij}^{\pm}(t)u_{j}^{\beta} \propto
\delta^{\alpha\beta}$.
The parity and eigenstate projected correlator, 
\begin{align}
 G^{\alpha}_{\pm}& \equiv v_{i}^{\alpha}G^{\pm}_{ij}(t)u_{j}^{\alpha} ,
 \label{projected_cf_final}
\end{align}
is then analyzed to obtain masses of energy-states. 

A eigenvector analysis of a symmetric matrix having orthogonal
eigenvectors can be constructed by inserting
${G(t_{0})}^{-\frac{1}{2}}\, {G(t_{0})}^{\frac{1}{2}}=I$, in
Eq.~(\ref{eq:right_evalue_eq}) and multiplying by
${G(t_{0})}^{\frac{1}{2}}$ from the left,
\begin{align}
{G(t_{0})}^{-\frac{1}{2}}\, G(t_{0}+\triangle t)\, {G(t_{0})}^{-\frac{1}{2}}\,
{G(t_{0})}^{\frac{1}{2}}\, u^{\alpha} & = c^{\alpha}\,
{G(t_{0})}^{\frac{1}{2}}u^{\alpha}\, ,
\label{eqn:symmetric_evalue_deriv} \\
{G(t_{0})}^{-\frac{1}{2}}\, G(t_{0}+\triangle t)\, {G(t_{0})}^{-\frac{1}{2}}\,
w^{\alpha} & = c^{\alpha}\, w^{\alpha} \, ,
\label{eqn:symmetric_evalue}
\end{align} 
where, $w^\alpha = {G(t_{0})}^{\frac{1}{2}}\, u^\alpha$ and
$[{G(t_{0})}^{-\frac{1}{2}}\, G(t_{0}+\triangle t)\,
  {G(t_{0})}^{-\frac{1}{2}}]$ is a real symmetric matrix, with
orthogonal eigenvectors ${w}^{\alpha}.$ The vector $u^\alpha$ may be
recovered from the $w^\alpha$ via $u^\alpha =
{G(t_{0})}^{-\frac{1}{2}}\, w^\alpha$.

The PACS-CS $2+1$ flavor dynamical-fermion
configurations~\cite{Aoki:2008sm} made available through the
ILDG~\cite{Beckett:2009cb} are used herein.  These configurations use
the non-perturbatively ${\cal{O}}(a)$-improved Wilson fermion action
and the Iwasaki-gauge action~\cite{Iwasaki:1983ck}. The lattice volume
is $32^{3}\times 64$, with $\beta=1.90$ providing a lattice spacing of
$a=0.0907$ fm and a physical volume of $\approx (2.90\, \rm{fm})^{3}$.
Five values of the (degenerate) up and down quark masses are
considered, with hopping parameter values of $\kappa_{ud}=0.13700,
0.13727, 0.13754, 0.13770\text{ and }0.13781$, corresponding to pion
masses of $m_{\pi}$ = 0.702, 0.572, 0.413, 0.293, 0.156
GeV~\cite{Aoki:2008sm}; for the strange quark $\kappa_{s}=0.13640$.
Gauge-invariant Gaussian smearing~\cite{Gusken:1989qx} is used at the
fermion source and sink with a fixed smearing fraction and four
different smearing levels including 16, 35, 100, and 200
sweeps~\cite{Mahbub:2010rm,Menadue:2011pd}.

The complete set of local interpolating fields with three different
spin-flavor combinations for the spin-$\frac{1}{2}$ nucleon are
considered herein,
\begin{align}
\label{eq:interp_x1}
\chi_1(x) &= \epsilon^{abc}\, (u^{Ta}(x)\, C{\gamma_5}\, d^b(x)\, )\, u^{c}(x)\, , \\
\label{eq:interp_x2}
\chi_2(x) &= \epsilon^{abc}\, (u^{Ta}(x)\, C\, d^b(x)\, )\, {\gamma_5}\, u^{c}(x)\, , \\
\label{eq:interp_x4}
\chi_4(x) &= \epsilon^{abc}\, (u^{Ta}(x)\, C{\gamma_5}{\gamma_4}\,
d^b(x)\, )\, u^{c}(x).
\end{align}
Each interpolator has a unique Dirac structure giving rise to
different spin-flavor combinations. Moreover, as each spinor has upper
and lower components, with the lower components containing an implicit
derivative, different combinations of zero, one and two-derivative
interpolators are provided.  The interpolator $\chi_{4}$ is the time
component of the local spin-$\frac{3}{2}$ isospin-$\frac{1}{2}$
interpolator which also couples to spin-$\frac{1}{2}$ states. It
provides a different linear combination of zero- and two-derivative
terms complementary to $\chi_1$.

\begin{figure}[tb]
  \begin{center} 
  \includegraphics[angle=90,height=0.30\textwidth]{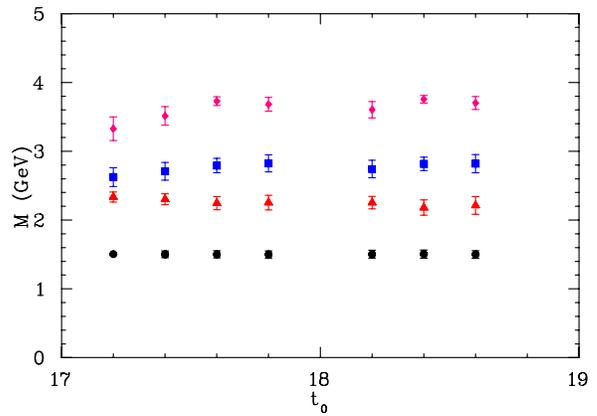}
 \caption{(Color online). $N{\frac{1}{2}}^{-}$ energy-states from a
   $4\times 4$ correlation matrix analysis of the $\chi_{1}$
   interpolator at the lightest pion mass of $m_{\pi}=156$ MeV. The
   variational parameters $t_{0}$ and $\triangle t$ are shown at the
   major and minor tick marks respectively. The LHS of the diagram
   refers to $t_{0}=17$, while the RHS is for $t_{0}=18.$}
\label{fig:m-_4x4.x1x1.t0deltat}
 \end{center}
\vspace{-0.4cm}
\end{figure}

\begin{figure}[tb]
  \begin{center} 
  \includegraphics[angle=90,height=0.30\textwidth]{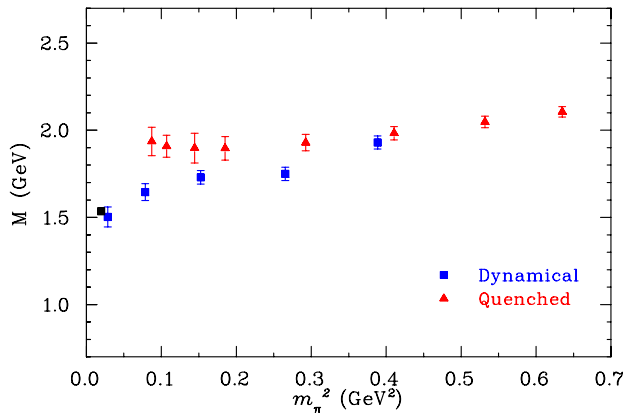}
 \caption{(Color online). Dynamical and quenched results for the
   lowest $N{\frac{1}{2}}^{-}$ energy-state using the $\chi_{1}$
   interpolator.}
 \label{fig:m.x1x1_Dyn_vs_Quench.negP}
 \end{center}
\vspace{-0.4cm}
\end{figure}

\begin{figure}[tb]
  \begin{center} 
  \includegraphics[angle=90,height=0.30\textwidth]{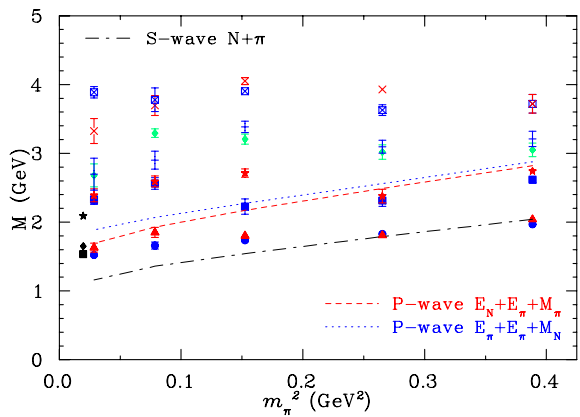}
 \caption{(Color online). $N{\frac{1}{2}}^{-}$ energy-states from an
   $8\times 8$ correlation matrix of $\chi_{1}$ and $\chi_{2}$
   interpolators, for the pion mass range of $156$ to $702$ MeV. The
   physical $N{\frac{1}{2}}^{-}$ spectrum~\cite{Nakamura:2010zzi} is
   shown at the far left.}
 \label{fig:m.x1x2.8x8.negP}
 \end{center}
\vspace{-0.4cm}
\end{figure}

In Fig.~\ref{fig:m-_4x4.x1x1.t0deltat}, projected masses of the
$N{\frac{1}{2}}^{-}$ states are presented from a $4 \times 4$
correlation matrix constructed from the interpolator $\chi_1$ and four
different smearing levels.  The dependence of the results on the
variational parameters $t_{0}$ and $\triangle t$ is illustrated.
While the lowest energy-state is almost independent of $t_{0}$ and
$\triangle t$, the excited states show some dependence at smaller
$t_{0}$ and $\triangle t$ values. The energy-states at
$(t_{0},\triangle t)=(18,2)$ provide the best balance between the
systematic and statistical uncertainties~\cite{Mahbub:2010rm} and
these parameters are therefore selected for our numerical study.

In Fig.~\ref{fig:m.x1x1_Dyn_vs_Quench.negP} we show results for the
lowest energy-state from dynamical and quenched~\cite{Mahbub:2010me}
QCD simulations.  As anticipated, the quenched and dynamical results
are in agreement in the heavy quark-mass region.  However, in the
light quark-mass regime the results are significantly different as the
effects of the light sea quarks become increasingly important.  Only
the dynamical results approach the physical value and this provides
strong evidence for the non-trivial role of light sea-quark degrees of
freedom to the structure of Nature's hadron spectrum.

To explore the nearby second energy state, $S_{11}\,(1650),$ we extend
our analysis to include the interpolators $\chi_{2}$ and $\chi_{4}$
with a variety of smearing levels.  The results of an $8 \times 8$
correlation-matrix analysis of $\chi_1$ and $\chi_2$ interpolators
with four levels of smearing are presented in
Fig.~\ref{fig:m.x1x2.8x8.negP}.

The flow of the lowest two energy states towards the physical values
is remarkable. The results at the two heaviest pion masses sit close
to the scattering S-wave $N+\pi$ threshold indicating that the results
for these heaviest pion masses may be scattering states.  However, in
the light quark-mass region these states move above the threshold.

It is likely that the three-quark interpolators used herein have
relatively small couplings to the scattering states at the light quark
masses relative to the states observed and are not resolved in the
correlation-matrix analysis.  Hence, a combination of five- and
three-quark interpolators in a correlation matrix analysis is highly
desirable for future investigations to ensure better overlap with the
multi-particle states.  This type of novel work using the stochastic
LapH method is in progress~\cite{Morningstar:2011ka}. It is because
the coupling to multi-particle states at light quark masses is heavily
suppressed, that it is meaningful to compare lattice results with the
central values of experimentally measured hadron resonances.

A similar situation prevails for the second pair of states in the
spectrum, where the states sit close to the $P$-wave
$E_{N}+E_{\pi}+M_{\pi}$ and $E_{\pi}+E_{\pi}+M_{N}$ threshold
scattering states with back-to-back momenta of one lattice unit, $\vec
p = (2 \pi/ L_x, 0, 0)$.  The apparent flow of these states in the
light-quark region toward the physical $S_{11}\,(2090)$ state is also
interesting.
 
In presenting the results of Fig.~\ref{fig:m.x1x2.8x8.negP} and
assigning symbols to each of the energy levels observed at a
particular quark mass, it is necessary to track the evolution of the
states from one quark mass to the next.  We have done this through a
consideration of the evolution of the eigenvectors as the quark mass
is changed.

Consider $M$ interpolating fields making an $M\times M$
parity-projected correlation matrix $G(t)$ and its associated
symmetric generalized eigenvalue equation of
Eq.~(\ref{eqn:symmetric_evalue}).  Using the normalization $\sum_{i}^M
\left \vert w_{i}^\alpha \right \vert^{2} = 1$, the quantity
$\vec{w}^{\alpha}(m_{q}) \cdot
\vec{w}^{\beta}(m_{q})=\delta_{\alpha\beta}$.
This feature enables the use of the generalized measure
\begin{align}
{\mathcal W}^{\alpha \beta}(m_q, m_{q^\prime}) & = \vec{w}^{\alpha}(m_{q}) \cdot \vec{w}^{\beta}(m_{q^\prime})
 \label{generalised_measure}
\end{align}
to identify the states most closely related as we move from quark mass
$m_{q}$ to an adjacent quark mass $m_{q^\prime}$.  The state numbers
$\alpha$ and $\beta$ are assigned in order of increasing projected
eigenstate energy at the quark masses $m_{q}$ and $m_{q^\prime}$
respectively.  Typical results for this generalized measure of
eigenvector overlap are presented in
Table~\ref{table:wdotw_k13754_k13770}.

For each value of state index $\alpha$ there is only one value of
$\beta$ where the magnitude of the entry is significantly larger than
all others and approaching unity.  The most relevant entries for
consideration are the immediate neighbors of $\alpha$ where a crossing
of the eigenvectors moves the largest entry off the diagonal.

This measure provides a clear identification of how states in the
spectrum at quark mass $m_q$ are associated with states at the next
value of quark mass, $m_{q^\prime}$.  For example, the results of
Table~\ref{table:wdotw_k13754_k13770} indicate the first four states
at $m_{q^\prime}$ appear with the same ordering in the spectrum as
observed at $m_q$, the fifth state at $m_{q^\prime}$ is associated
with the sixth state at $m_q$ and vice versa and similarly for the
seventh and eighth states.  We note that while the central values of
the energies have changed order, the error bars are sufficiently large
that one cannot conclude that an avoided energy level crossing has
taken place in moving from quark mass $m_q$ to $m_{q^\prime}$.

\begin{table}
 \begin{center}
\caption{The scalar product $\vec{w}^{\alpha}(m_{q}) \cdot
  \vec{w}^{\beta}(m_{q^\prime})$ for $\kappa = 0.13754$
  ($m_{\pi}=413\,\rm{MeV}$) and $\kappa^\prime = 0.13770$
  ($m_{\pi}=293\,\rm{MeV}$) for an $8\times 8$ correlation matrix of
  $\chi_{1}$ and $\chi_{2}$ with four different levels of smearing.
  State numbers $\alpha$ and $\beta$ correspond to row and column number,
  respectively.}
 \label{table:wdotw_k13754_k13770}
 \begin{tabular}{cccccccc}
 \hline
  \textbf{0.91} & 0.40 & 0.02 & 0.02 & 0.01 & -0.05 & 0.00 & 0.00 \\
  0.40 & \textbf{-0.91} & 0.00 & 0.01 & -0.02 & 0.01 & -0.01 & 0.00 \\
  -0.01 & -0.01 & \textbf{0.96} & -0.27 & 0.01 & -0.01 & 0.00 & 0.02 \\
  -0.03 & 0.00 & 0.27 & \textbf{0.96} & 0.01 & 0.01 & 0.02 & 0.00 \\
  0.04 & 0.03 & 0.01 & -0.01 & -0.22 & \textbf{0.97} & 0.02 & 0.01 \\
  0.01 & -0.01 & -0.01 & -0.01 & \textbf{0.98} & 0.22 & 0.04 & 0.00 \\
  0.00 & 0.00 & -0.02 & 0.01 & 0.01 & -0.01 & -0.12 & \textbf{0.99} \\
  0.01 & -0.01 & 0.00 & -0.02 & -0.04 & -0.03 & \textbf{0.99} & 0.12 \\
  \hline
\end{tabular}
 \vspace{-0.4cm} 
\end{center}
\end{table}

\begin{figure}[tb]
  \begin{center} 
  \includegraphics[angle=90,height=0.26\textwidth]{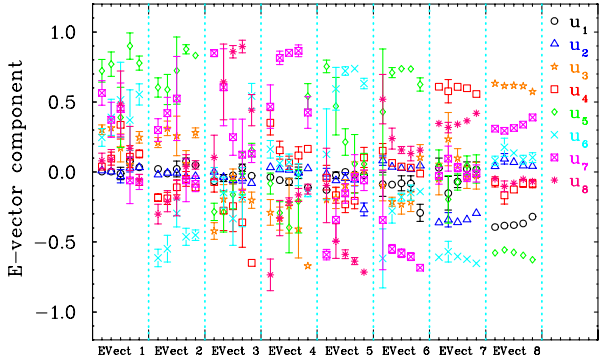}
 \caption{(Color online). The components of the eigenvector $u^\alpha$
   providing the amplitude for each interpolating field at the source
   for creating the state $\alpha$. The states are labeled by the
   eigenvector (EVect) number with the ordering as provided in
   Fig.~\ref{fig:m.x1x2.8x8.negP} at the heaviest quark mass. For each
   EVect, the eigenvector components are plotted from left to right in
   order of increasing quark mass.  In the legend, $(u_{1},\, u_{2})$,
   $(u_{3},\, u_{4})$, $(u_{5},\, u_{6})$ and $(u_{7},\, u_{8})$
   correspond to the smearing-sweep levels of $16$, $35$, $100$ and
   $200$ respectively.  Odd numbers in the subscripts correspond to
   the contribution from the $\chi_{1}$ interpolator, whereas, even
   numbers correspond to $\chi_{2}$.}
 \label{fig:evectors_8x8_x1x2.negP}
 \end{center}
\vspace{-0.4cm}
\end{figure}

The components of the eigenvector $u^\alpha$, providing the amplitude
for each interpolating field at the source for creating the state
$\alpha$, are provided in Fig.~\ref{fig:evectors_8x8_x1x2.negP}.  A
non-trivial contribution from both the $\chi_{1}$ and $\chi_{2}$
interpolators for the lowest two energy-states is evident. The
scalar-diquark interpolator $\chi_{1}$ dominates the lowest
energy-state.  On the other hand, both $\chi_{1}$ and $\chi_{2}$
interpolators have large contributions to the second energy state
where their strengths appear with opposite signs.  The eigenvector
components typically display a slow evolution as the quark mass is
changed.

\begin{figure}[t]
  \begin{center} 
  \includegraphics[angle=90,height=0.30\textwidth]{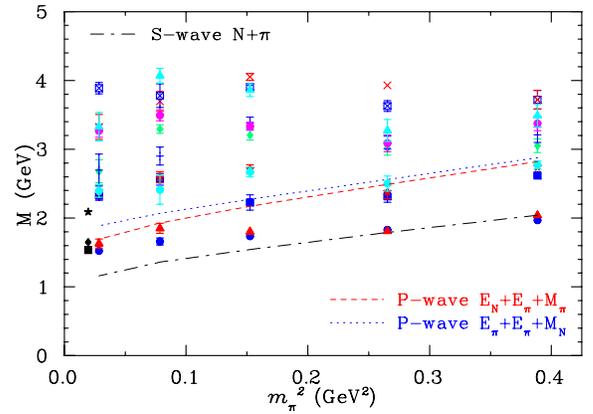}
 \caption{(Color online). Masses of 12 low-lying $N{\frac{1}{2}}^{-}$
   energy states from two $8\times 8$ correlation matrices of
   $\chi_{1},\chi_{2}$ and $\chi_{1},\chi_{4}$.}
 \label{fig:m-.12x12}
\end{center}
\vspace{-0.4cm}
\end{figure}

The energy-states for our complete analysis are presented in
Fig.~\ref{fig:m-.12x12}. The results are drawn from two $8 \times 8$
correlation-matrix analyses for pairs of $\chi_{1},\chi_{2}$ and
$\chi_{1},\chi_{4}$.  The matrices are formed with each interpolator
having four levels of smearing. Whereas the $\chi_{1}$, $\chi_{2}$ and
$\chi_{2}$, $\chi_{4}$ analyses reveal a similar spectrum, four new
states are revealed in the $\chi_{1}$, $\chi_{4}$ analysis providing
the resolution of 12 low-lying states in our analysis.

In the quark model based on $SU(6)$ spin-flavor symmetry, the odd
parity $(1535)$ and $(1650)$ states belong to the negative parity,
$L=1$, 70-plet representation of $SU(6)$.  As three spin-$\frac{1}{2}$
quarks may combine to a total spin of $s=\frac{1}{2}$ or
$\frac{3}{2}$, the $L=1$ state can couple two different ways to
provide a $J=\frac{1}{2}$ state, hence providing two orthogonal
spin-$\frac{1}{2}$ states in the $L=1$, 70-plet representation. Both
of these states have a width of $\approx 150$ MeV. The lowest two
energy states revealed here are similarly close in mass, as
illustrated in Fig.~\ref{fig:m-.12x12}, in accord with the $SU(6)$
quark model.

\begin{figure}[tb]
  \begin{center} 
  \includegraphics[angle=90,height=0.30\textwidth]{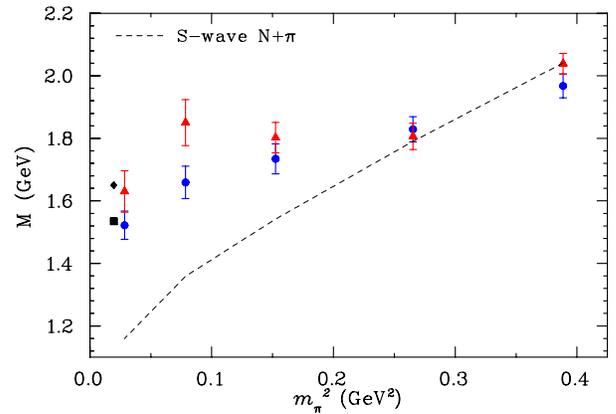}
 \caption{(Color online). The quark mass dependence of the lowest two
   lowest-lying $N{\frac{1}{2}}^{-}$ states are compared with the
   $S$-wave scattering threshold.}
 \label{fig:m.x1x2.2states.negP.swave}
 \end{center}
\vspace{-0.4cm}
\end{figure}

These two lowest-lying $N{\frac{1}{2}}^{-}$ states are presented in
Fig.~\ref{fig:m.x1x2.2states.negP.swave} in comparison with the
$S$-wave scattering threshold.  These lattice results, providing the
first examination of the odd-parity nucleon spectrum at a pion mass as
low as 156 MeV, display remarkable agreement with the physical values.
They represent a significant achievement for lattice regularized QCD
in describing Nature.

Although both these low-lying states are quite similar at the two
heaviest quark masses, their approach to the physical values in the
light quark-mass region are different.  Significant chiral curvature
is evident, in particular for the second state.  It will be
interesting to explore the mass dependence of these states using
effective field theory techniques and to repeat these studies on
matched lattices of different volume when they become available.
Future studies will endeavor to observe the multi-particle scattering
states and determine the resonance parameters of these states from the
first principles of QCD.

This research was undertaken on the NCI National Facility in Canberra,
Australia, which is supported by the Australian Commonwealth
Government. We also acknowledge eResearch SA for generous grants of
supercomputing time.  This research is supported by the Australian
Research Council.


\begin{thebibliography}{16}%
\makeatletter
\providecommand \@ifxundefined [1]{%
 \@ifx{#1\undefined}
}%
\providecommand \@ifnum [1]{%
 \ifnum #1\expandafter \@firstoftwo
 \else \expandafter \@secondoftwo
 \fi
}%
\providecommand \@ifx [1]{%
 \ifx #1\expandafter \@firstoftwo
 \else \expandafter \@secondoftwo
 \fi
}%
\providecommand \natexlab [1]{#1}%
\providecommand \enquote  [1]{``#1''}%
\providecommand \bibnamefont  [1]{#1}%
\providecommand \bibfnamefont [1]{#1}%
\providecommand \citenamefont [1]{#1}%
\providecommand \href@noop [0]{\@secondoftwo}%
\providecommand \href [0]{\begingroup \@sanitize@url \@href}%
\providecommand \@href[1]{\@@startlink{#1}\@@href}%
\providecommand \@@href[1]{\endgroup#1\@@endlink}%
\providecommand \@sanitize@url [0]{\catcode `\\12\catcode `\$12\catcode
  `\&12\catcode `\#12\catcode `\^12\catcode `\_12\catcode `\%12\relax}%
\providecommand \@@startlink[1]{}%
\providecommand \@@endlink[0]{}%
\providecommand \url  [0]{\begingroup\@sanitize@url \@url }%
\providecommand \@url [1]{\endgroup\@href {#1}{\urlprefix }}%
\providecommand \urlprefix  [0]{URL }%
\providecommand \Eprint [0]{\href }%
\@ifxundefined \urlstyle {%
  \providecommand \doi  [0]{\begingroup \@sanitize@url \@doi}%
  \providecommand \@doi [1]{\endgroup \@@startlink {\doibase
  #1}doi:\discretionary {}{}{}#1\@@endlink }%
}{%
  \providecommand \doi  [0]{doi:\discretionary{}{}{}\begingroup
  \urlstyle{rm}\Url }%
}%
\providecommand \doibase [0]{http://dx.doi.org/}%
\providecommand \Doi [0]{\begingroup \@sanitize@url \@Doi }%
\providecommand \@Doi  [1]{\endgroup\@@startlink{\doibase#1}\@@Doi}%
\providecommand \@@Doi [1]{#1\@@endlink}%
\providecommand \selectlanguage [0]{\@gobble}%
\providecommand \bibinfo  [0]{\@secondoftwo}%
\providecommand \bibfield  [0]{\@secondoftwo}%
\providecommand \translation [1]{[#1]}%
\providecommand \BibitemOpen [0]{}%
\providecommand \bibitemStop [0]{}%
\providecommand \bibitemNoStop [0]{.\EOS\space}%
\providecommand \EOS [0]{\spacefactor3000\relax}%
\providecommand \BibitemShut  [1]{\csname bibitem#1\endcsname}%
\bibitem [{\citenamefont {Nakamura}\ \emph {et~al.}(2010)\citenamefont
  {Nakamura} \emph {et~al.}}]{Nakamura:2010zzi}%
  \BibitemOpen
  \bibfield  {author} {\bibinfo {author} {\bibfnamefont {K.}~\bibnamefont
  {Nakamura}} \emph {et~al.} (\bibinfo {collaboration} {Particle Data Group}),\
  }\Doi {10.1088/0954-3899/37/7A/075021} {\bibfield  {journal} {\bibinfo
  {journal} {J. Phys.},\ }\textbf {\bibinfo {volume} {G37}},\ \bibinfo {pages}
  {075021} (\bibinfo {year} {2010})}\BibitemShut {NoStop}%
\bibitem [{\citenamefont {Bulava}\ \emph {et~al.}(2009)\citenamefont {Bulava}
  \emph {et~al.}}]{Bulava:2009jb}%
  \BibitemOpen
  \bibfield  {author} {\bibinfo {author} {\bibfnamefont {J.~M.}\ \bibnamefont
  {Bulava}} \emph {et~al.},\ }\Doi {10.1103/PhysRevD.79.034505} {\bibfield
  {journal} {\bibinfo  {journal} {Phys. Rev.},\ }\textbf {\bibinfo {volume}
  {D79}},\ \bibinfo {pages} {034505} (\bibinfo {year} {2009})},\ \Eprint
  {http://arxiv.org/abs/0901.0027} {arXiv:0901.0027 [hep-lat]} \BibitemShut
  {NoStop}%
\bibitem [{\citenamefont {Bulava}\ \emph {et~al.}(2010)\citenamefont {Bulava}
  \emph {et~al.}}]{Bulava:2010yg}%
  \BibitemOpen
  \bibfield  {author} {\bibinfo {author} {\bibfnamefont {J.}~\bibnamefont
  {Bulava}} \emph {et~al.},\ }\Doi {10.1103/PhysRevD.82.014507} {\bibfield
  {journal} {\bibinfo  {journal} {Phys. Rev.},\ }\textbf {\bibinfo {volume}
  {D82}},\ \bibinfo {pages} {014507} (\bibinfo {year} {2010})},\ \Eprint
  {http://arxiv.org/abs/1004.5072} {arXiv:1004.5072 [hep-lat]} \BibitemShut
  {NoStop}%
\bibitem [{\citenamefont {Engel}\ \emph {et~al.}(2010)\citenamefont {Engel},
  \citenamefont {Lang}, \citenamefont {Limmer}, \citenamefont {Mohler},\ and\
  \citenamefont {Schafer}}]{Engel:2010my}%
  \BibitemOpen
  \bibfield  {author} {\bibinfo {author} {\bibfnamefont {G.~P.}\ \bibnamefont
  {Engel}}, \bibinfo {author} {\bibfnamefont {C.~B.}\ \bibnamefont {Lang}},
  \bibinfo {author} {\bibfnamefont {M.}~\bibnamefont {Limmer}}, \bibinfo
  {author} {\bibfnamefont {D.}~\bibnamefont {Mohler}}, \ and\ \bibinfo {author}
  {\bibfnamefont {A.}~\bibnamefont {Schafer}} (\bibinfo {collaboration} {BGR
  [Bern-Graz-Regensburg]}),\ }\Doi {10.1103/PhysRevD.82.034505} {\bibfield
  {journal} {\bibinfo  {journal} {Phys. Rev.},\ }\textbf {\bibinfo {volume}
  {D82}},\ \bibinfo {pages} {034505} (\bibinfo {year} {2010})},\ \Eprint
  {http://arxiv.org/abs/1005.1748} {arXiv:1005.1748 [hep-lat]} \BibitemShut
  {NoStop}%
\bibitem [{\citenamefont {Edwards}\ \emph {et~al.}(2011)\citenamefont
  {Edwards}, \citenamefont {Dudek}, \citenamefont {Richards},\ and\
  \citenamefont {Wallace}}]{Edwards:2011jj}%
  \BibitemOpen
  \bibfield  {author} {\bibinfo {author} {\bibfnamefont {R.~G.}\ \bibnamefont
  {Edwards}}, \bibinfo {author} {\bibfnamefont {J.~J.}\ \bibnamefont {Dudek}},
  \bibinfo {author} {\bibfnamefont {D.~G.}\ \bibnamefont {Richards}}, \ and\
  \bibinfo {author} {\bibfnamefont {S.~J.}\ \bibnamefont {Wallace}},\ }\Doi
  {10.1103/PhysRevD.84.074508} {\bibfield  {journal} {\bibinfo  {journal}
  {Phys. Rev.},\ }\textbf {\bibinfo {volume} {D84}},\ \bibinfo {pages} {074508}
  (\bibinfo {year} {2011})},\ \Eprint {http://arxiv.org/abs/1104.5152}
  {arXiv:1104.5152 [hep-ph]} \BibitemShut {NoStop}%
\bibitem [{\citenamefont {Mahbub}\ \emph {et~al.}(2012)\citenamefont {Mahbub},
  \citenamefont {Kamleh}, \citenamefont {Leinweber}, \citenamefont {Moran},\
  and\ \citenamefont {Williams}}]{Mahbub:2010rm}%
  \BibitemOpen
  \bibfield  {author} {\bibinfo {author} {\bibfnamefont {M.~S.}\ \bibnamefont
  {Mahbub}}, \bibinfo {author} {\bibfnamefont {W.}~\bibnamefont {Kamleh}},
  \bibinfo {author} {\bibfnamefont {D.~B.}\ \bibnamefont {Leinweber}}, \bibinfo
  {author} {\bibfnamefont {P.~J.}\ \bibnamefont {Moran}}, \ and\ \bibinfo
  {author} {\bibfnamefont {A.~G.}\ \bibnamefont {Williams}} (\bibinfo
  {collaboration} {CSSM Lattice}),\ }\href@noop {} {\bibfield  {journal}
  {\bibinfo  {journal} {Phys. Lett.},\ }\textbf {\bibinfo {volume} {B707}},\
  \bibinfo {pages} {389} (\bibinfo {year} {2012})},\ \Eprint
  {http://arxiv.org/abs/1011.5724} {arXiv:1011.5724 [hep-lat]} \BibitemShut
  {NoStop}%
\bibitem [{\citenamefont {Menadue}\ \emph {et~al.}(2012)\citenamefont
  {Menadue}, \citenamefont {Kamleh}, \citenamefont {Leinweber},\ and\
  \citenamefont {Mahbub}}]{Menadue:2011pd}%
  \BibitemOpen
  \bibfield  {author} {\bibinfo {author} {\bibfnamefont {B.~J.}\ \bibnamefont
  {Menadue}}, \bibinfo {author} {\bibfnamefont {W.}~\bibnamefont {Kamleh}},
  \bibinfo {author} {\bibfnamefont {D.~B.}\ \bibnamefont {Leinweber}}, \ and\
  \bibinfo {author} {\bibfnamefont {M.~S.}\ \bibnamefont {Mahbub}},\ }\Doi
  {10.1103/PhysRevLett.108.112001} {\bibfield  {journal} {\bibinfo  {journal}
  {Phys. Rev. Lett.},\ }\textbf {\bibinfo {volume} {108}},\ \bibinfo {pages}
  {112001} (\bibinfo {year} {2012})},\ \Eprint {http://arxiv.org/abs/1109.6716}
  {arXiv:1109.6716 [hep-lat]} \BibitemShut {NoStop}%
\bibitem [{\citenamefont {Michael}(1985)}]{Michael:1985ne}%
  \BibitemOpen
  \bibfield  {author} {\bibinfo {author} {\bibfnamefont {C.}~\bibnamefont
  {Michael}},\ }\href@noop {} {\bibfield  {journal} {\bibinfo  {journal} {Nucl.
  Phys.},\ }\textbf {\bibinfo {volume} {B259}},\ \bibinfo {pages} {58}
  (\bibinfo {year} {1985})}\BibitemShut {NoStop}%
\bibitem [{\citenamefont {Luscher}\ and\ \citenamefont
  {Wolff}(1990)}]{Luscher:1990ck}%
  \BibitemOpen
  \bibfield  {author} {\bibinfo {author} {\bibfnamefont {M.}~\bibnamefont
  {Luscher}}\ and\ \bibinfo {author} {\bibfnamefont {U.}~\bibnamefont
  {Wolff}},\ }\href@noop {} {\bibfield  {journal} {\bibinfo  {journal} {Nucl.
  Phys.},\ }\textbf {\bibinfo {volume} {B339}},\ \bibinfo {pages} {222}
  (\bibinfo {year} {1990})}\BibitemShut {NoStop}%
\bibitem [{\citenamefont {Mahbub}\ \emph {et~al.}(2009)\citenamefont {Mahbub}
  \emph {et~al.}}]{Mahbub:2009aa}%
  \BibitemOpen
  \bibfield  {author} {\bibinfo {author} {\bibfnamefont {M.~S.}\ \bibnamefont
  {Mahbub}} \emph {et~al.},\ }\Doi {10.1016/j.physletb.2009.07.063} {\bibfield
  {journal} {\bibinfo  {journal} {Phys. Lett.},\ }\textbf {\bibinfo {volume}
  {B679}},\ \bibinfo {pages} {418} (\bibinfo {year} {2009})},\ \Eprint
  {http://arxiv.org/abs/0906.5433} {arXiv:0906.5433 [hep-lat]} \BibitemShut
  {NoStop}%
\bibitem [{\citenamefont {Aoki}\ \emph {et~al.}(2009)\citenamefont {Aoki} \emph
  {et~al.}}]{Aoki:2008sm}%
  \BibitemOpen
  \bibfield  {author} {\bibinfo {author} {\bibfnamefont {S.}~\bibnamefont
  {Aoki}} \emph {et~al.} (\bibinfo {collaboration} {PACS-CS}),\ }\href@noop {}
  {\bibfield  {journal} {\bibinfo  {journal} {Phys. Rev. D},\ }\textbf
  {\bibinfo {volume} {79}},\ \bibinfo {pages} {034503} (\bibinfo {year}
  {2009})},\ \Eprint {http://arxiv.org/abs/0807.1661} {arXiv:0807.1661
  [hep-lat]} \BibitemShut {NoStop}%
\bibitem [{\citenamefont {Beckett}\ \emph {et~al.}(2011)\citenamefont {Beckett}
  \emph {et~al.}}]{Beckett:2009cb}%
  \BibitemOpen
  \bibfield  {author} {\bibinfo {author} {\bibfnamefont {M.~G.}\ \bibnamefont
  {Beckett}} \emph {et~al.},\ }\Doi {10.1016/j.cpc.2011.01.027} {\bibfield
  {journal} {\bibinfo  {journal} {Comput. Phys. Commun.},\ }\textbf {\bibinfo
  {volume} {182}},\ \bibinfo {pages} {1208} (\bibinfo {year} {2011})},\ \Eprint
  {http://arxiv.org/abs/0910.1692} {arXiv:0910.1692 [hep-lat]} \BibitemShut
  {NoStop}%
\bibitem [{\citenamefont {Iwasaki}(1983)}]{Iwasaki:1983ck}%
  \BibitemOpen
  \bibfield  {author} {\bibinfo {author} {\bibfnamefont {Y.}~\bibnamefont
  {Iwasaki}},\ }\href@noop {} { (\bibinfo {year} {1983})},\ \bibinfo {note}
  {uTHEP-118}\BibitemShut {NoStop}%
\bibitem [{\citenamefont {Gusken}(1990)}]{Gusken:1989qx}%
  \BibitemOpen
  \bibfield  {author} {\bibinfo {author} {\bibfnamefont {S.}~\bibnamefont
  {Gusken}},\ }\href@noop {} {\bibfield  {journal} {\bibinfo  {journal} {Nucl.
  Phys. Proc. Suppl.},\ }\textbf {\bibinfo {volume} {17}},\ \bibinfo {pages}
  {361} (\bibinfo {year} {1990})}\BibitemShut {NoStop}%
\bibitem [{\citenamefont {Mahbub}\ \emph {et~al.}(2010)\citenamefont {Mahbub},
  \citenamefont {Kamleh}, \citenamefont {Leinweber}, \citenamefont {Cais},\
  and\ \citenamefont {Williams}}]{Mahbub:2010me}%
  \BibitemOpen
  \bibfield  {author} {\bibinfo {author} {\bibfnamefont {M.~S.}\ \bibnamefont
  {Mahbub}}, \bibinfo {author} {\bibfnamefont {W.}~\bibnamefont {Kamleh}},
  \bibinfo {author} {\bibfnamefont {D.~B.}\ \bibnamefont {Leinweber}}, \bibinfo
  {author} {\bibfnamefont {A.~O.}\ \bibnamefont {Cais}}, \ and\ \bibinfo
  {author} {\bibfnamefont {A.~G.}\ \bibnamefont {Williams}},\ }\Doi
  {10.1016/j.physletb.2010.08.049} {\bibfield  {journal} {\bibinfo  {journal}
  {Phys. Lett.},\ }\textbf {\bibinfo {volume} {B693}},\ \bibinfo {pages} {351}
  (\bibinfo {year} {2010})},\ \Eprint {http://arxiv.org/abs/1007.4871}
  {arXiv:1007.4871 [hep-lat]} \BibitemShut {NoStop}%
\bibitem [{\citenamefont {Morningstar}\ \emph {et~al.}(2011)\citenamefont
  {Morningstar} \emph {et~al.}}]{Morningstar:2011ka}%
  \BibitemOpen
  \bibfield  {author} {\bibinfo {author} {\bibfnamefont {C.}~\bibnamefont
  {Morningstar}} \emph {et~al.},\ }\Doi {10.1103/PhysRevD.83.114505} {\bibfield
   {journal} {\bibinfo  {journal} {Phys. Rev.},\ }\textbf {\bibinfo {volume}
  {D83}},\ \bibinfo {pages} {114505} (\bibinfo {year} {2011})},\ \Eprint
  {http://arxiv.org/abs/1104.3870} {arXiv:1104.3870 [hep-lat]} \BibitemShut
  {NoStop}%
\end{thebibliography}
\end{document}